\documentclass[aps,pra,twocolumn,superscriptaddress,showpacs]{revtex4-1}

\usepackage{amsmath}
\usepackage{graphicx}
\usepackage{bm}
\usepackage[T1]{fontenc}
\usepackage[latin1]{inputenc}
\usepackage{times}
\usepackage{amssymb}

\begin{document}

\title{Deterministic light focusing in space and time through multiple scattering media with a Time-Resolved Transmission Matrix approach}

\author{Mickael Mounaix} \email{mickael.mounaix@lkb.ens.fr} \affiliation{Laboratoire Kastler Brossel, ENS-PSL Research University, CNRS, UPMC Paris Sorbonne, Coll\`{e}ge de France, 24 rue Lhomond, 75005 Paris, France } 
\author{Hugo Defienne} \affiliation{Laboratoire Kastler Brossel, ENS-PSL Research University, CNRS, UPMC Paris Sorbonne, Coll\`{e}ge de France, 24 rue Lhomond, 75005 Paris, France } \affiliation{Department of Electrical Engineering, Princeton University, Princeton, NJ, 08544, USA}
\author{Sylvain Gigan}  \affiliation{Laboratoire Kastler Brossel, ENS-PSL Research University, CNRS, UPMC Paris Sorbonne, Coll\`{e}ge de France, 24 rue Lhomond, 75005 Paris, France }

\begin{abstract}
We report a method to characterize the propagation of an ultrashort pulse of light through a multiple scattering medium by measuring its time-resolved transmission matrix. This method is based on the use of a spatial light modulator together with a coherent time-gated detection of the transmitted speckle field. Using this matrix, we demonstrate the focusing of the scattered pulse at any arbitrary position in space and time after the medium. Our approach opens new perspectives for both fundamental studies and applications in imaging and coherent control in disordered media .
\end{abstract}


\maketitle

When coherent light propagates in a scattering medium, light exits the system under the form of a speckle pattern~\cite{goodman1976some}, the result of complex interference effects between all the scattered waves. All the information carried by the incoming light are scrambled in the transmitted field. Multiple scattering of light is then an adverse effect for most optical imaging applications~\cite{ntziachristos2010going}. This wave mixing process is certainly complex, but it is deterministic. In the last years, wavefront shaping techniques have exploited the deterministic nature of light scattering to control light propagation in scattering media, using Spatial Light Modulators (SLMs). These techniques have demonstrated the control of coherent light propagating through a layer of paint or through multimode fibers using either an iterative optimization approach~\cite{vellekoop2007focusing} or digital phase conjugation~\cite{papadopoulos2012focusing}. 
Wavefront shaping technniques have also been used to measure the monochromatic scattering matrix of a disordered medium~\cite{popoff2010measuring}. This complex operator links any input field of the medium to its corresponding output field and can be used to focus light or image through a disordered medium~\cite{popoff2010measuring,popoff2010image,choi2011overcoming}. The matrix approach has been applied to many different systems, such as multimode fibers~\cite{carpenter2014110x110} and extended to different fields of optics, such as photoacoustics~\cite{chaigne2014controlling} or quantum optics~\cite{defienne2016two}.

Light propagation in a disordered system has also been investigated with low-coherence sources such as ultra-short pulse lasers. In this situation, light is mixed spatially but also stretched temporally during its propagation and speckle patterns in space and time are generated at the output.  Pulse recompression can be achieved via spectral shaping~\cite{mccabe_spatio-temporal_2011}.  Counter intuitively, shaping only the spatial properties also enables spatio-temporal control thanks to the spatio-temporal coupling performed by the medium, as demonstrated in acoustics~\cite{lemoult2009manipulating} and in optics~\cite{aulbach_control_2011,katz2011focusing,morales2015delivery,paudel2013focusing,mounaix2015spatiotemporal}. Other approaches based on the measurement of a time-resolved reflection matrix have also been proposed for focusing~\cite{PhysRevLett.111.243901} or imaging ~\cite{kang2015imaging,badon2015smart} at a target depth inside a scattering medium. In this regime, the time-gated detection of back-scattered photons aims at selecting a certain depth of the scattering sample, essentially by selecting ballistic photons,  similarly to optical coherence tomography. However, when light propagates through a disordered medium with an optical thickness larger of several transport mean free path, the diffusive regime is reached in transmission and no ballistic photons can be detected at the output. Recently, the deterministic control of an ultra-short pulse of light propagating through a multiple-scattering medium has been achieved by measuring its multi-spectral transmission matrix (MSTM)~\cite{mounaix2015spatiotemporal,andreoli2015deterministic}. This matrix characterizes light propagation for all the different wavelengths that compose the incoming pulse and enables a deterministic spatio-temporal control of the scattered pulse at the output by exploiting the time-frequency duality. This approach has nevertheless an important practical drawback in that it requires the full knowledge of the spectral information (i.e the full measurement of the MSTM) to control accurately the output temporal speckle. Indeed, most of the information content of the MSTM is superfluous if one is only interested in manipulating light at a specific arrival time at the output. 

In this letter, we report the first experimental measurement of a time-resolved transmission matrix (TRTM) of a scattering medium in the diffusive regime using a coherent time-gated detection. Unlike the MSTM approach, we demonstrate that a TRTM measured for a given single time enables an efficient spatio-temporal focusing of the pulse at at the chosen arrival time after the medium. Finally, we show that the full knowledge of the TRTM enables shaping more sophisticated spatio-temporal profiles of the pulse at the output, such as pump-probe profiles. 

\begin{figure}
\centering
\includegraphics[scale=1]{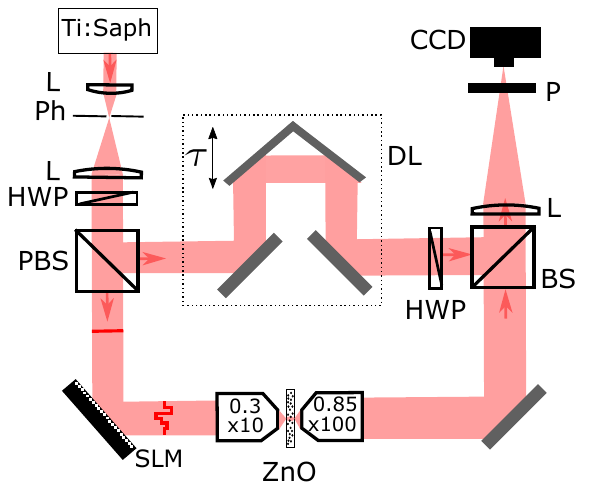}
\caption{\label{manip} (Color online) Apparatus for measuring a time resolved transmission matrix (TRTM) of a scattering medium. An ultra-short pulse of light generated by a Ti:Sapphire laser (central frequency: 800nm ; $\tau_0 \approx 110$ fs FWHM) is split between a reference and a control path using a combination of a half-wave plate (HWP) and a polarizing beam splitter (PBS). In the control arm, the pulse is reflected by a SLM and injected into a sample of ZnO nanoparticles with an thickness of approximately 100 $\mu m$ using a microscope objective. The scattered light is collected in transmission by another objective and interfere  with the reference pulse onto a beam splitter (BS). The microscope objective together with the lens (L) image the output surface of the scattering medium onto a charge-coupled device (CCD) camera. A polarizer (P) is used to select only one output polarization. The delay line (DL) inserted in the reference path controls the relative optical path delay between the reference and the scattered pulse. (Ph): Pinhole}
\end{figure}

Fig.~\ref{manip} sketches the experimental setup used to measure the TRTM of a scattering sample. A Ti:Sapphire laser source (MaiTai, Spectra Physics) produces an ultra-short pulse centered around 800 nm with a duration of $\tau_0 \approx 110$ fs. The pulse is split between a reference and a control arm using an half-wave plate (HWP) and a polarizing beam splitter (PBS). In the control arm, a phase-only SLM (LCOS-SLM, Hamamatsu X10468) modulates the wavefront of the reflected pulse. The shaped pulse is then injected in a thick layer of ZnO nanoparticles (thickness of approximately 100 $\mu m$) using a microscope objective. Scattered light is collected on the other side of the medium using a microscope objective.  Another microscope objective and a lens (L) are used to collect the scattered light and image the output surface of the medium onto a charge-coupled device (CCD) camera (Allied Vision, Manta G-046). Scattered and reference pulses are recombined before the camera using a beam-splitter (BS) and one polarization is selected using a linear polarizer (P). A delay line (DL) inserted in the reference arm sets the relative optical path delay between the reference and the scattered pulses.

An ultra-short pulse of light propagating in a multiple-scattering medium follows a large distribution of optical diffusive paths determined by the exact position of the scatterers. All these optical paths interfere at the output and generate a complex spatio-temporal speckle pattern~\cite{mosk2012controlling}. Spatial and temporal features of the speckle are characterized respectively by the size of a speckle grain and by the traversal time of the medium, related to its dwell time $\tau_m$. The dwell time depends only of the medium properties and refers to the duration during which the photons stay confined inside the medium, due to the broad path length distribution in the sample because of multiple scattering events ~\cite{Curry:11, Patterson:89}. The spatio-temporal mixing performed by the medium is the equivalent of the spatio-spectral coupling that has been investigated and reported in previous works ~\cite{mccabe_spatio-temporal_2011, mounaix2015spatiotemporal, andreoli2015deterministic}. One can define the spectral bandwidth of the medium $\Delta \lambda_m$~\cite{mounaix2015spatiotemporal,andreoli2015deterministic}, which  is the minimal difference in input wavelengths required to generate uncorrelated speckle patterns at the output, and which only depends on the medium properties. The spectral bandwidth  is inversely proportional to the dwell time $\tau_m$ of the photons. Indeed, a pulse with a spectral bandwidth larger than the spectral bandwidth of the medium- or in an equivalent way a pulse with a temporal duration smaller than the dwell time of the medium - will experience spectral dispersion - or temporal broadening - during propagation.

\begin{figure}
\centering
\includegraphics[scale=0.8]{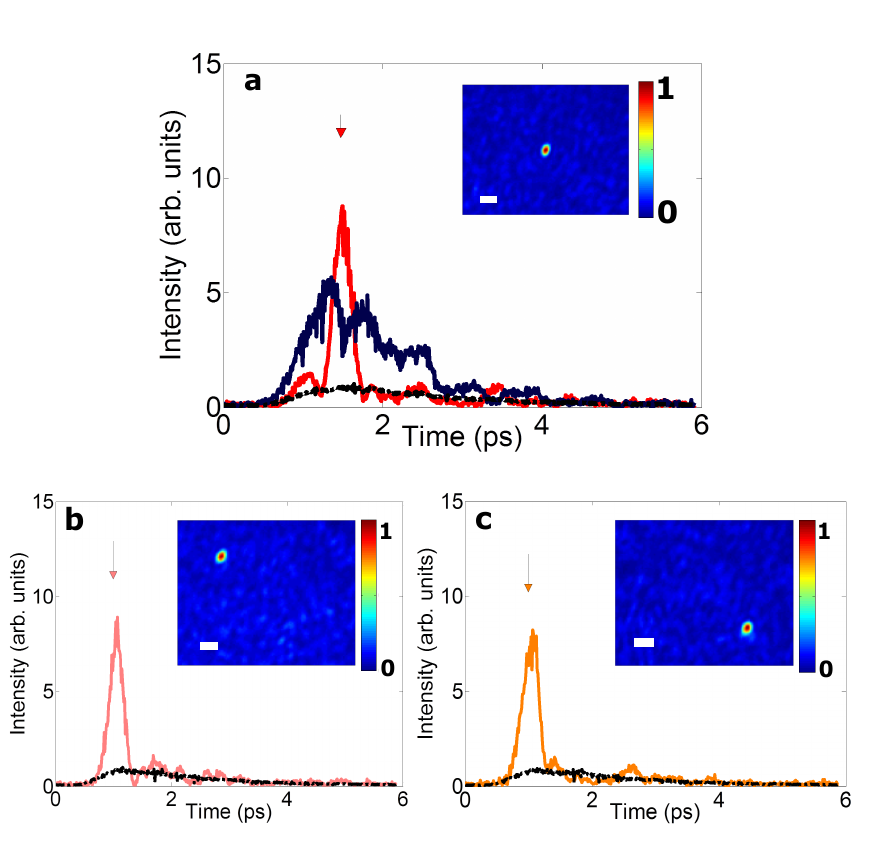}
\caption{\label{STfoc} (Color online) Results of spatio-temporal focusing at a given output position and a given time by phase-conjugating the time-resolved transmission matrix $H(t_a)$ measured at the arrival time $t_a = 1.6$ ps (denoted by the top arrows). (a)  Temporal profiles recorded at the targeted spatial output position in cases of spatio-temporal focusing (red) and spatial-only focusing (blue line) processes. The spatial-only focusing process is performed by applying the phase-conjugation technique to a monochromatic transmission matrix measured at the central wavelength of the incoming pulse. The target position is visible on the CCD image in inset. An average output temporal profile of the pulse (black line) is obtained by averaging temporal profiles over 100 different speckle grains at the output. The dwell time of the medium is evaluated from this averaged profile to be about $\tau_m \simeq 2$ ps. (b) and (c) show temporal profiles and the corresponding CCD camera images for spatio-temporal focusing processes performed at two different spatial positions using same transmission matrix $H(t_a)$. Averaged temporal profiles are drawn in black. Scale bars on the CCD images in inset correspond to 2 $\mu m$.}
\end{figure}

Because a CCD camera is not fast enough to resolve the temporal structure of the pulse at the output, we estimate $\tau_m$ using Interferometric Cross Correlation (ICC) technique~\cite{monmayrant2010newcomer}. This linear pulse characterization technique enables to retrieve the temporal envelop of the output speckle by interfering it with the Fourier limited pulse of the reference arm. For this purpose, a set of intensity images is recorded with the CCD camera while the delay line of the reference arm is scanned. The signal measured at each pixel of the camera is then an interferogram produced by the combination of the output speckle measured at this specific pixel and the reference pulse. The temporal envelop of the output pulses are finally retrieved from their interferograms by applying a low pass filter. As presented on Fig.~\ref{STfoc}a, the temporal structure of the output pulse averaged over 100 different speckle grains shows the expected exponential decay profile~\cite{mounaix2015spatiotemporal}. The dwell time is estimated from the averaged profile to be about $\tau_m \simeq 2 $ ps $ \sim 20 \tau_0 $. The output pulse is therefore broadened on average by a factor $\simeq 20$ relative to the original pulse.

Thanks to the linearity of the scattering process and the stability of the medium, the propagation of an optical pulse can be described using a scattering matrix formalism. The optical field measured at a given arrival time $t_a$ at the output is linked to the input field by the formula: 

\begin{equation} \label{eqTM}
E^{\text{out}} (t_a) =H(t_a) E^{\text{in}}
\end{equation}
where $E^{\text{in}}$ is a complex vector containing amplitude and phase values of the field for each input mode (i.e a SLM pixel), $E^{\text{out}} (t_a)$ is a complex vector containing amplitude and phase values of the field for each output mode (i.e a CCD camera pixel) measured at a specific arrival time $t_a$, and  $H(t_a)$ is the transmission matrix measured at $t_a$ connecting inputs to outputs. A complete set of matrices $\{ H(t_a) \}_{t_a}$ forms the full TRTM of the scattering medium. 

Spatio-temporal focusing of the transmitted pulse is achieved first by measuring the time-resolved matrix $H(t_a)$. The matrix is measured column by column by recording the output fields for a set of $N$ SLM patterns at the input. Each transmitted field is retrieved from intensity measurements on the CCD camera at the output using a phase-stepping holographic process, as in a monochromatic case~\cite{popoff2010measuring}. However, in contrast with~\cite{popoff2010measuring}, the ultrashort reference pulse provides a time-gating since the interference can only come from a time-window given by the pulse duration. Furthermore, the targeted detection time $t_a$ can be set by adjusting the delay line. 

\begin{figure}
\centering
\includegraphics[scale=1]{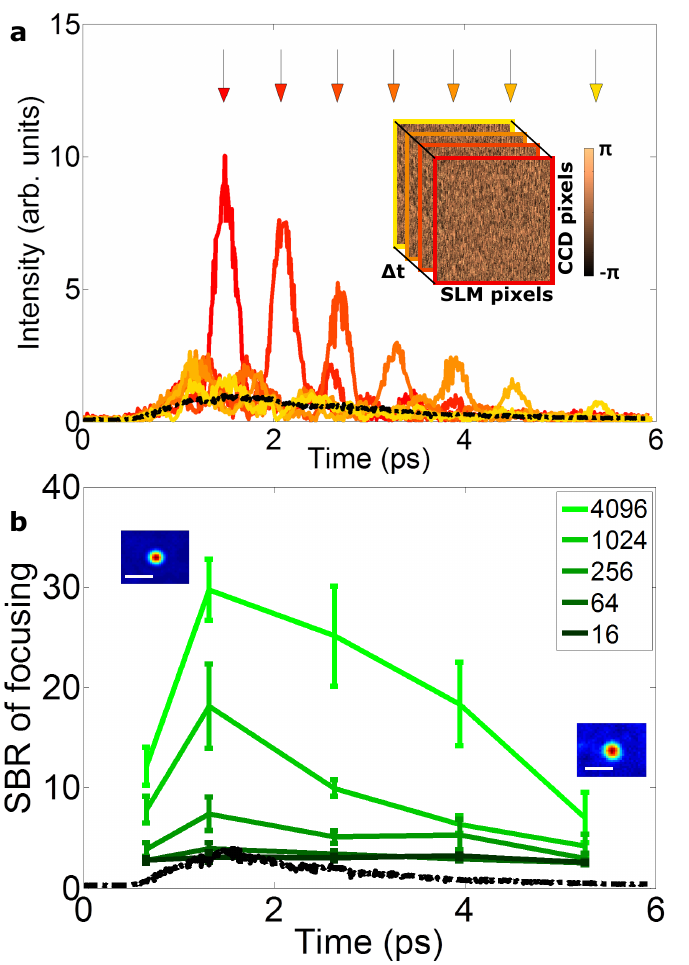}
\caption{\label{multi_time} (Color online) Spatio-temporal control of light with a set of transmission matrices $\{ H(t_a) \}_{t_a}$ measured at different arrival times. (a)  Temporal profiles acquired using spatio-temporal focusing processes at the same output position and different arrival times (colored arrows). The spatially averaged speckle is used as a reference temporal profile (black curve). Phase components of the TRTM measured are drawn in inset (b) Signal-to-background ratio (SBR) measured on the CCD camera for different targeted arrival time and different number of modes controlled at the input. The SBR is the ratio of the intensity at the targeted pixel of the CCD camera over the mean intensity value of the full speckle image. Scale bar is 2$\mu m$.}
\end{figure}

The pattern to be programmed on the SLM for focusing light in space and time is then calculated using a phase conjugation approach as in Ref.~\cite{popoff2010measuring}: 

\begin{equation} \label{phase_conjugation}
E^{\text{in}}=H^{\dagger}(t_a) E^{\text{target}}_x
\end{equation}
where $ H^{\dagger}(t_a)$ is the conjugate transpose of $H(t_a)$ and $E^{\text{target}}_x$ is a null vector with a coefficient 1 at the row corresponding to the targeted position $x$ on the camera. Performing a digital phase conjugation (see Eq.~\ref{phase_conjugation})  corresponds to controlling the phases of the input modes - and then consequently the global phases of the speckle patterns they generate at the output. Thus we precisely set these speckles to interfere constructively at the targeted output mode~\cite{popoff2010measuring}. Thanks to the time gating due to the reference pulse during the matrix measurement, this constructive interference process occurs only for a specific arrival time. This operator acts precisely as a time reversal operator for the arrival time $t_a$ and light is focused both in space and time at the output~\cite{derode1995robust}.

Results of spatio-temporal focusing of a $110$ fs pulse transmitted through a thick layer of paint using a set of $N=256$ input modes are shown on Fig.~\ref{STfoc}. The time-resolved matrix $H(t_a)$ is measured at the arrival time $t_a = 1.6$ ps as indicated by the top arrows. For a single time, the matrix measurement process takes about 2 minutes,  this time being mostly limited by the refresh rate of the SLM. The SLM is then programmed using the phase conjugation approach and the temporal profile of the output pulse at the targeted spatial position is reconstructed with an ICC measurement. As presented on Fig.~\ref{STfoc}a, the temporal profile of the resulting focused pulse (red) shows a peak of intensity centered at $t_a$ with a temporal width of $150$ fs, close to the width of the Fourier-limited incoming pulse ($110$ fs). This pulse profile is compared to the temporal profile (blue) acquired using a spatial-only focusing process, which is achieved by phase conjugation of the monochromatic transmission matrix measured at the central wavelength of the pulse~\cite{mounaix2015spatiotemporal}. As expected, no temporal compression is observed in this case, but the intensity measured at the targeted pixel on the CCD camera remains more intense than the background~\cite{mounaix2015spatiotemporal}. By changing $E^{\text{target}}$ in Eq.~\ref{phase_conjugation}, the output pulse can be focused at any arbitrary output positions. As presented on Fig.~\ref{STfoc}b and Fig.~\ref{STfoc}c, the resulting temporal profiles for focusing at two different spatial positions show the same temporal compression at the arrival time $t_a = 1.6$ ps. As expected, the intensity enhancement observed on the CCD camera are also similar (insets).

Full spatio-temporal control of the pulse at the output requires the measurement of a larger set of time-resolved transmission matrices.  On Fig.~\ref{multi_time}, 7 transmission matrices have been recorded at 7 different arrival times distributed between $t_1 = 1.6$ ps and $t_7 = 4.6$ ps. The time gap between two targeted arrival times $|t_i-t_j|$ is set larger than the time-width of a temporal speckle grain, which corresponds also to the time-width of the initial input pulse~\cite{mounaix2015spatiotemporal}, to ensure the matrices are well uncorrelated with each other. As presented on Fig.~\ref{multi_time}a, each matrix of the set can be independently used to perform spatio-temporal focusing at the different arrival times. 

Efficiency of the spatio-temporal focusing process can be analyzed by measuring a t signal-to-background ratio (SBR) on the CCD images recorded at the output. The SBR is the ratio of the intensity at the targeted pixel of the CCD camera over the mean intensity value of the full speckle image, integrated over the acquisition time of the camera. As shown on Fig.~\ref{multi_time}b, SBR values depend both of the targeted arrival time and the number of modes controlled at the input. For a given arrival time, the SBR increases with the number of mode controlled, and for a fixed number of controlled modes, maximal values of SBR are always reached for the targeted arrival time that corresponds to the maximum of intensity in the averaged temporal profile. Besides, we observe that spatio-temporal focusing may be achieved even for very long arrival time ($t_7 \approx 4.6$ ps) where the photon rate is very low, provided the number of modes controlled is sufficiently high. As the input light is broadband, values of SBR measured in this experiment are much lower than the one observed in the monochromatic case~\cite{mosk2012controlling}, and also much lower than the peak to background intensity that would be obtained at the target time $t_a$. 

\begin{figure}
\centering
\includegraphics[scale=1]{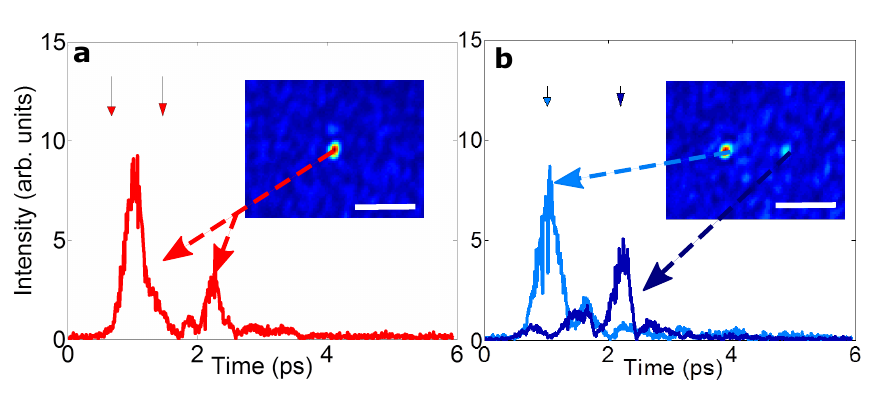}
\caption{\label{deux_tgtm} (Color online) Complex spatio-temporal shaping of the output pulse by exploiting the full TRTM. (a) Temporal profiles recorded using a spatio-temporal focusing process at one spatial position and two different arrival times $t_1 = 1$ ps and $t_2 = 2.2$ ps. CCD image in inset shows the position of the spatial target. (b) Temporal profiles recorded using a spatio-temporal focusing process at the same two different arrival times and at two different spatial positions, visible on the CCD image in inset. Scale bar is 5$\mu m$.}
\end{figure}

The TRTM can also be used to shape more complex spatio-temporal profiles at the output. For example, spatio-temporal focusing at two different positions $x_1$ and $x_2$ and two different times $t_1$ and $t_2$ simultaneously is achieved by calculating the SLM pattern using a combination of the corresponding matrices $H(t_1)$ and $H(t_2)$:
\begin{equation}
E^{\text{in}}=(H^{\dagger}(t_1) E^\text{target}_{x_1} + H^{\dagger}(t_2) E^\text{target}_{x_2} )
\end{equation}
where $E^\text{target}_{x_1} [ E^\text{target}_{x_2} ]$ is a null vector containing a coefficient 1 at the line that corresponds to the position $x_1[x_2]$ of the camera. Fig.~\ref{deux_tgtm}a shows results of spatio-temporal focusing in the particular case of $x_1=x_2$ with $t_1 = 1$ ps and $t_2 = 2.2$ ps. Logically, we observe that the intensity values of the two successive peaks are approximately half the intensities obtained for the same arrival times using simple spatio-temporal focusing processes presented in Fig.~\ref{multi_time} with the same number of input modes ($N=256$). Such a deterministic pump-probe-like pulse has an interesting potential for applications in light-matter interaction in scattering media~\cite{sapienza2011long}. Results obtained by generalizing this process to the case $x_1 \neq x_2$ are shown on Fig.~\ref{deux_tgtm}b. The output speckle is now focused simultaneously at two different times ($t_1 = 1$ ps and $t_2 = 2.2$ ps, visible on temporal profiles) and at two different speckle grains (visible on the CCD image in inset).

It is interesting to compare the TRTM approach demonstrated here to the multispectral approach of Ref~\cite{mounaix2015spatiotemporal}. Although the MSTM and the TRTM contains in principle the same information, the most adapted depend on the experiment to be performed. While the MSTM could address the spectral dispersion by a phase control on all the spectral components of the output pulse, the TRTM enables a direct temporal refocusing by adjusting the optical paths with the same arrival time at the output of the medium. Clearly, re-compressing the pulse at the output at a given time is much easier and faster using the TRTM, as it requires measuring a single time-resolved transmission matrix, rather than measuring the MSTM for all the spectral components then recombining them accordingly. Nonetheless, narrowband focusing as well as more refined  pulse  control, in phase and in amplitude, of the output pulse is more straightforward using the MSTM. 

In conclusion, we have demonstrated deterministic spatio-temporal focusing of an output pulse after propagation through a disordered medium, with a single measurement of a transmission matrix of the medium at a given arrival time. Combining several scattering matrices from the full TRTM enable spatio-temporal focusing at different time using a single SLM. This approach, complementary to the other spectral or temporal approach to light control in complex media,  could enable potential applications in multiphotonic imaging and light-matter interactions in disordered media.

\begin{acknowledgments}
The author would like to thanks Thomas Chaigne, Samuel Gr\'{e}sillon and Ori Katz for fruitful discussions.
This work was funded by the European Research Council (grant no. 278025). S. G. is a member of the Institut Universitaire de France. 
\end{acknowledgments}


\begin{thebibliography}{99}
\bibitem{goodman1976some} J. W. Goodman, {\it JOSA} {\bf 66} 1145 (1976).
\bibitem{ntziachristos2010going} V. Ntziachristos,  {\it Nat. Meth.} {\bf 7} 603 (2010).
\bibitem{vellekoop2007focusing} I. M. Vellekoop and A. P. Mosk,  {\it Opt. Lett.} {\bf 32} 2309--2311 (2007).
\bibitem{papadopoulos2012focusing} I. N. Papadopoulos, S. Farahi, C. Moser and D. Psaltis ,  {\it Opt. Exp.} {\bf 20} 10583 (2012).
\bibitem{popoff2010measuring} S. M. Popoff  G. Lerosey, R. Carminati, M. Fink, A. C. Boccara, and S. Gigan, \prl{ 104} 100601 (2010).
\bibitem{popoff2010image} S. M. Popoff, G. Lerosey, M. Fink, A. C. Boccara, and S. Gigan, {\it Nat. Comm.} {\bf 1} 81 (2010).
\bibitem{choi2011overcoming} Y. Choi, T. D. Yang, C. Fang-Yen, P. Kang, K. J. Lee, R. R. Dasari, M. S. Feld,  and W. Choi, \prl{ 107} 023902 (2011).
\bibitem{carpenter2014110x110} J. Carpenter, B. J. Eggleton, and J. Schr{\"o}der,, {\it Opt. Exp.}  {\bf 22}, 96 (2014)
\bibitem{chaigne2014controlling} T. Chaigne, O. Katz, A. C. Boccara, M. Fink, E. Bossy,  and S. Gigan,  {\it Nat. Photon.} {\bf 8} 58 (2014).
\bibitem{defienne2016two} H. Defienne, M. Barbieri, I. A. Walmsley, B. J. Smith,  and S. Gigan {\it Sci. Adv.} {\bf 2} e1501054 (2016).
\bibitem{mccabe_spatio-temporal_2011}D. J. McCabe, A. Tajalli, D. R. Austin, P. Bondareff, I. A. Walmsley, S. Gigan,  and B. Chatel, {\it Nat. Comm.} {\bf 2} 447 (2011).
\bibitem{lemoult2009manipulating} F. Lemoult, G. Lerosey, J. de Rosny and M. Fink,  \prl{ 103} 173902 (2009).
\bibitem{aulbach_control_2011} J. Aulbach, B. Gjonaj, P. M. Johnson, A. P. Mosk,  and A. Lagendijk,  \prl{ 106} 103901 (2011) 
\bibitem{katz2011focusing} O. Katz, E. Small, Y. Bromberg, and Y. Silberberg,  {\it Nat. Photon.} {\bf 5} 372 (2011) 
\bibitem{morales2015delivery} E.  E.  Morales-Delgado,  S.  Farahi,  I.  N.  Papadopoulos, D. Psaltis,  and C. Moser,  {\it Opt. Exp.} {\bf 23} 9109 (2015).
\bibitem{paudel2013focusing} H. P. Paudel, C. Stockbridge, J. Mertz, and T. Bifano,   {\it Opt. Exp.}  {\bf 21}, 17299 (2013) 
\bibitem{mounaix2015spatiotemporal} M. Mounaix, D. Andreoli, H. Defienne, G. Volpe, O. Katz, S.  Gr\'{e}sillon,   and  S.  Gigan,   \prl{ 116} 253901 (2016) 
\bibitem{PhysRevLett.111.243901} Y. Choi, T. R. Hillman, W. Choi, N. Lue, R. R. Dasari, P. T. C. So, W. Choi,  and Z. Yaqoob,  \prl{ 111} 243901 (2013).
\bibitem{kang2015imaging} S. Kang, S. Jeong, W. Choi, H. Ko, T. D. Yang, J. H. Joo, J.-S. Lee, Y.-S. Lim, Q.-H. Park,  and W. Choi,  {\it Nat. Photon.} {\bf 9} 253 (2015)
\bibitem{badon2015smart} A. Badon, D. Li, G. Lerosey, A. C. Boccara, M. Fink,  and A. Aubry, {\it arXiv preprint} arXiv:1510.08613 (2015)
\bibitem{andreoli2015deterministic}D. Andreoli, G. Volpe, S. Popoff, O. Katz, S. Gr\'{e}sillon,  and S. Gigan,  {\it Sci. Rep.} {\bf 5} (2015).
\bibitem{mosk2012controlling}A. P. Mosk, A. Lagendijk, G. Lerosey, and M. Fink,  {\it Nat. Photon.} {\bf 6} 283--292 (2012) 
\bibitem{Curry:11} N.  Curry,  P.  Bondareff,  M.  Leclercq,  N.  F.  van  Hulst, R. Sapienza, S. Gigan,  and S.  Gr\'{e}sillon, {\it Opt. Lett.} {\bf 36} 3332--3334 (2011).
\bibitem{Patterson:89} M. S. Patterson, B. Chance, and B. C. Wilson, {\it Appl. Opt.} {\bf 28} 2331-2336 (1989) 
\bibitem{monmayrant2010newcomer} A. Monmayrant, S. Weber, and B. Chatel,  {\it Journal of Physics B: Atomic, Molecular and Optical Physics} {\bf 43} 103001 (2010).
\bibitem{derode1995robust} A. Derode, P. Roux, and M. Fink,   \prl{ 75} 4206 (1995) 
 \bibitem{sapienza2011long} R. Sapienza  P. Bondareff, R. Pierrat, B. Habert, R. Carminati, and N. Van Hulst,  \prl{ 106} 163902 (2011) 
 \end{thebibliography}
\end{document}